\documentclass{article}

\usepackage{arxiv}
\usepackage[T1]{fontenc}
\usepackage[utf8]{inputenc}
\usepackage{amsmath}
\usepackage{amssymb}
\usepackage{booktabs}
\usepackage{graphicx}
\usepackage{hyperref}
\usepackage[numbers]{natbib}
\usepackage{url}

\title{citecheck: An MCP Server for Automated Bibliographic Verification and Repair in Scholarly Manuscripts}

\author{
  Junhyeok Lee\\
  \href{https://orcid.org/0000-0001-7489-5829}{ORCID: 0000-0001-7489-5829}\\
  \texttt{evening0619@gmail.com}
}

\hypersetup{
  colorlinks=true,
  linkcolor=blue,
  citecolor=blue,
  urlcolor=blue
}

\begin{document}
\maketitle

\begin{abstract}
Reference lists in scholarly manuscripts frequently contain errors, including incorrect identifiers, incomplete metadata, misattributed authors, and mismatches between preprint and published versions. These problems are tedious to repair manually and have become more visible in workflows that rely on large language models, which can fabricate or corrupt citations. We present \textit{citecheck}, a TypeScript system and MCP server for automated bibliographic verification and repair in paper-like project folders. Given a manuscript file or workspace, citecheck selects the most likely paper artifact, extracts references from \texttt{.bib}, \texttt{.tex}, \texttt{.md}, \texttt{.txt}, or \texttt{.docx}, validates entries against PubMed, Crossref, arXiv, and Semantic Scholar, and returns structured correction proposals together with replacement-safety diagnostics. The current repository provides a working research prototype with multi-pass retrieval, manifestation-aware matching, policy-gated rewrite planning, and 47 passing tests covering repair behavior, malformed payload handling, transport failures, and MCP exposure. We position citecheck as infrastructure for agentic scholarly editing and as a practical guardrail against both traditional reference errors and LLM-induced citation hallucinations.
\end{abstract}


\section{Introduction}
Reference lists are one of the least automation-friendly parts of scholarly writing. Manuscripts often contain incomplete entries, missing DOIs, inconsistent author formatting, and mismatches between preprints and their later journal or conference manifestations. Manual verification is slow because authors must cross-reference several external registries entry by entry.

This long-standing problem has become more urgent in workflows that use large language models (LLMs). Citation hallucinations are now well documented: models can fabricate references, blend metadata from different papers, and emit plausible-looking but invalid identifiers \citep{walters2023,linardon2025}. Recent large-scale studies suggest that the problem is already visible in published and preprint literature, especially in fast-moving AI venues \citep{xu2026,ansari2026}. Those reports are recent and some remain preprints, so their exact rates should be interpreted cautiously, but the direction of the problem is clear.

Reference inaccuracies were already common before LLM-assisted writing. Prior studies found substantial error rates across public health, social work, and medical literature \citep{eichorn1987,spivey2004,mogull2017}. What has changed is the operational context: authors and AI agents increasingly need a tool that can verify a manuscript's existing bibliography directly, not merely format or store curated library entries.

We present \textit{citecheck}, an MCP-native bibliography verification and repair system \citep{mcp_spec}. Rather than expecting users to prepare a clean bibliography file in advance, citecheck accepts a project path, identifies the most paper-like source file, extracts references, validates them against external scholarly metadata services, and returns both corrected output and explicit diagnostics. The system is intentionally conservative. It treats bibliography repair as a safety problem as much as a retrieval problem, distinguishing between review-only findings and overwrite-safe replacements.

The main contributions of this work are:
\begin{itemize}
  \item an MCP workflow that exposes workspace scanning, reference analysis, rewrite planning, and controlled application as separate tools;
  \item a multi-source, manifestation-aware verification pipeline spanning PubMed, Crossref, arXiv, and Semantic Scholar \citep{pubmed_api,crossref_api,arxiv_api,semantic_scholar_api};
  \item a policy layer that blocks unsafe replacements and surfaces curation worklists rather than silently rewriting ambiguous entries;
  \item an arXiv-oriented software paper that documents the current system, its repository-backed evidence, and its limitations.
\end{itemize}

\section{Related Work and Motivation}
Existing reference-management tools such as Zotero, Mendeley, and EndNote focus on organizing and formatting user-curated libraries \citep{zotero,mendeley,endnote}. They are indispensable for managing references, but they do not independently verify whether a manuscript's current reference list matches authoritative external records.

Other tools cover adjacent parts of the problem. Parsers such as AnyStyle and GROBID extract structured metadata from raw citation strings or documents \citep{anystyle,grobid}. Single-purpose utilities such as \texttt{doi2bib} resolve individual identifiers to formatted BibTeX entries \citep{doi2bib}. Metadata APIs such as Crossref, PubMed, arXiv, and Semantic Scholar expose authoritative or enrichment-oriented records \citep{crossref_api,pubmed_api,arxiv_api,semantic_scholar_api}. However, none of these components alone addresses the end-to-end task of selecting manuscript files, extracting references, cross-validating against multiple sources, classifying discrepancies, and producing replacement-safe outputs for agent workflows.

This gap matters for two audiences. Researchers need a practical final-pass verification step before submission. Developers of AI-assisted writing environments need a programmatic backend that can be invoked by agents rather than a GUI-first reference manager. citecheck is aimed at both groups: it behaves as a machine-consumable verification service but remains useful on ordinary paper files.

\section{System Overview}
citecheck is designed around project folders rather than isolated citation strings. The top-level workflow is:
\begin{enumerate}
  \item scan a workspace or inspect a single file path;
  \item select the most likely manuscript or bibliography artifact;
  \item extract the references section and normalize entries;
  \item query external metadata sources;
  \item cluster candidate matches and resolve preferred manifestations;
  \item apply replacement-safety policy checks;
  \item emit structured review results and optional rewrite patches.
\end{enumerate}

The MCP server exposes this workflow through six tools: \texttt{scan\_workspace}, \texttt{analyze\_references}, \texttt{plan\_reference\_rewrite}, \texttt{apply\_reference\_rewrite}, \texttt{repair\_paper}, and \texttt{citecheck\_version}. Separating analysis from write-back is deliberate. In agent settings, those should not be conflated.

\begin{figure}[t]
  \centering
  \includegraphics[width=\textwidth]{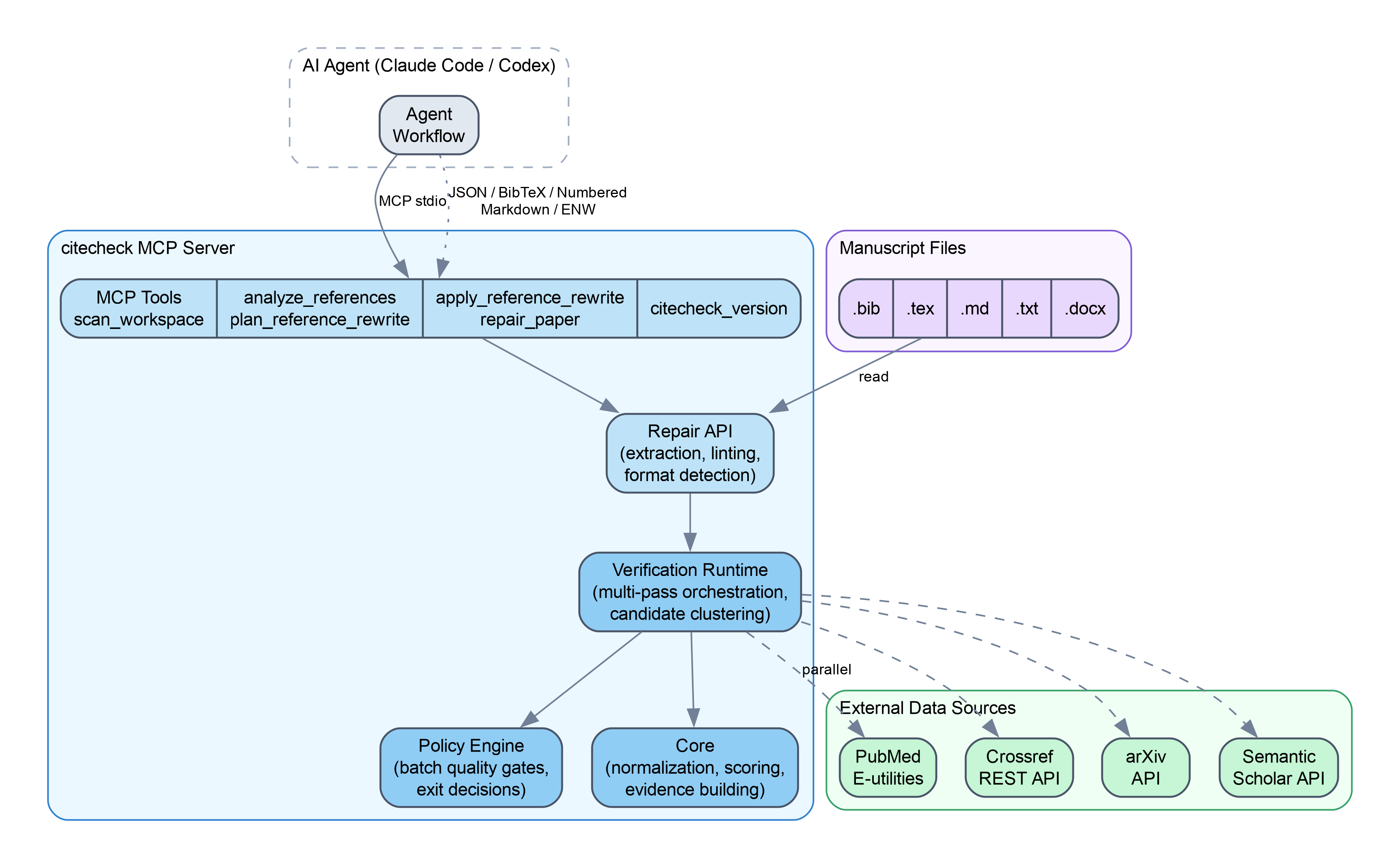}
  \caption{Simplified architecture of citecheck. The MCP server delegates to the repair API, the runtime orchestrates multi-pass verification across external connectors, and the policy engine evaluates replacement safety before write-back.}
  \label{fig:architecture}
\end{figure}

\begin{table}[t]
\caption{Supported inputs and outputs in the current implementation.}
\label{tab:io}
\centering
\begin{tabular}{ll}
\toprule
Category & Current support \\
\midrule
Input files & \texttt{.bib}, \texttt{.tex}, \texttt{.md}, \texttt{.txt}, \texttt{.docx} \\
Modes & \texttt{review}, \texttt{replacement} \\
Output formats & JSON, BibTeX, numbered text, Markdown, EndNote \\
Write modes & preview, sidecar, replace \\
Sources & PubMed, Crossref, arXiv, Semantic Scholar \\
\bottomrule
\end{tabular}
\end{table}

\section{Method}
\subsection{Document Selection and Extraction}
Workspace selection uses file-extension priority, filename hints such as \texttt{reference}, \texttt{paper}, \texttt{manuscript}, and \texttt{draft}, and content heuristics. Hidden or generated directories are skipped. For plain-text manuscripts, citecheck detects section headers such as ``References,'' ``Bibliography,'' and ``Works Cited.'' For \texttt{.tex} files, it traces bibliography resources referenced by commands such as \texttt{\textbackslash bibliography} and \texttt{\textbackslash addbibresource}. For \texttt{.docx} files, the current implementation extracts paragraph text from the document XML.

\subsection{Multi-Source Verification}
No single bibliographic registry covers all relevant literature. PubMed is authoritative for biomedical records. Crossref provides broad DOI-centered coverage. arXiv supplies preprint metadata. Semantic Scholar provides additional enrichment, including citation-related metadata. citecheck queries enabled sources in parallel so that a single weak source does not necessarily force an unresolved result.

\subsection{Multi-Pass Retrieval and Matching}
Extracted entries are normalized into internal reference-input types and then evaluated through up to three retrieval passes. The first pass uses the most direct query. Later passes reformulate searches when evidence remains weak. Retrieved candidates are deduplicated, clustered, and compared against the normalized query using title, author, venue, year, and identifier signals. The system then derives issues such as title mismatch, identifier conflict, year disagreement, or insufficient evidence.

\subsection{Manifestation-Aware Resolution}
One underlying work may appear as a journal article, conference paper, or preprint. citecheck explicitly models this distinction and applies the preference policy ``journal $>$ conference $>$ arXiv'' when linked manifestations are available. Crucially, manifestation disagreements are not silently rewritten away. If the evidence supports a review-only state rather than a safe replacement, the entry remains flagged for curation.

\subsection{Policy-Gated Rewrite Planning}
Automated bibliography replacement is risky because a plausible but incorrect match can silently damage a manuscript. citecheck therefore separates analysis from modification. The policy engine inspects batch-level ratios of \texttt{not\_checked}, \texttt{unresolved}, \texttt{verified}, and \texttt{needs\_review} entries, together with failure classes such as transport, authentication, rate-limit, and payload-shape failures. Presets named \texttt{default}, \texttt{strict}, and \texttt{lenient} translate those signals into exit decisions.

This design keeps safety metadata first-class. A result may be useful for review while still being blocked for in-place replacement because of duplicate keys, unsafe citation-key rewrites, or manifestation conflicts. Rewrite planning therefore produces patch previews before any sidecar or in-place write is attempted.

\section{Implementation}
The current repository is a TypeScript workspace with approximately 6.8k lines across application and library code. The MCP stdio server wraps the repair API and exposes the tool surface directly to clients such as Codex or Claude Code. The runtime factory supports both live HTTP connectors and fixture-backed replay, allowing deterministic regression tests without requiring every test to hit external APIs.

At the API layer, citecheck returns more than corrected bibliography text. It also returns per-entry statuses, confidence information, evidence traces, bibliography lint findings, curation worklists, duplicate-key detection, source-health summaries, failure summaries, replacement status, and explicit key-mapping risk. This structured output is central to agent integration because it allows a caller to distinguish between ``best available suggestion'' and ``safe to apply automatically.''

\section{Current Evidence}
This repository does not yet provide a large benchmark over thousands of references, and we do not claim that it does. Instead, we report what is directly supported by the current codebase.

\begin{table}[t]
\caption{Automated test coverage visible in the repository at the time of writing.}
\label{tab:tests}
\centering
\begin{tabular}{lr}
\toprule
Test group & Count \\
\midrule
Repair pipeline tests & 17 \\
Connector and HTTP behavior tests & 8 \\
Runtime, factory, policy, and core tests & 10 \\
MCP integration tests & 3 \\
Evaluation and regression fixture tests & 9 \\
\midrule
Total & 47 \\
\bottomrule
\end{tabular}
\end{table}

The test suite covers end-to-end repair on BibTeX, Markdown, \LaTeX{}, and DOCX-style inputs; replacement planning and sidecar rewrite behavior; duplicate-key blocking; manifestation conflicts; malformed source payload classification; transport and rate-limit failures; and MCP tool helper behavior. The repository also contains small fixture-based evaluation checks, including an offline biomedical DOI-recovery regression. These fixtures are useful for regression protection but should not be mistaken for a broad empirical benchmark.

In internal use, citecheck has been applied to real manuscript preparation as a practical verification aid. We treat that experience as qualitative evidence of utility rather than as a formal evaluation result. The repository currently demonstrates engineering completeness and safety-oriented behavior more strongly than benchmark scale.

\section{Discussion and Limitations}
citecheck sits in a specific systems niche. It is neither a full reference manager nor a general literature-search engine. Its role is to bridge messy manuscript artifacts and the structured outputs needed by agentic editing workflows. This framing explains two core design choices: first, reference repair is treated as a file-and-workspace problem rather than a string-only problem; second, safety diagnostics are elevated alongside candidate retrieval.

The system also has clear limitations. Evaluation remains small-scale. The manifestation policy is heuristic rather than graph-complete. Semantic Scholar is optional and disabled by default in the runtime factory. Extraction is robust for several paper-like formats but does not attempt full document-layout understanding. We also have not yet reported user studies or large comparative benchmarks against alternative tooling.

Promising next steps include expanding fixture-backed evaluations, building a realistic benchmark of malformed bibliographies, measuring citation-key rewrite accuracy on real \LaTeX{} projects, and studying the tradeoff between review-only and replacement-safe operating modes in downstream author workflows.

\section{Conclusion}
We presented citecheck, an MCP server for automated bibliographic verification and repair in scholarly manuscripts and paper-like project folders. The system combines workspace-aware extraction, multi-source metadata retrieval, manifestation-aware matching, and policy-gated rewrite planning. Its current repository already provides a functional prototype with structured outputs and automated tests, while its limitations remain explicit. We view citecheck as infrastructure for dependable scholarly editing systems and as a practical safeguard against both traditional reference errors and newer LLM-driven citation failures.

\bibliographystyle{plainnat}
\bibliography{references}

\end{document}